# Adaptive Dual-Layer Web Application Firewall (ADL-WAF) Leveraging Machine Learning for Enhanced Anomaly and Threat Detection

Ahmed Sameh[1] and Sahar Selim[1]

Center for Informatics Science (CIS), School of Information Technology and Computer Science, Nile University, 26th of July Corridor, Sheikh Zayed City 12588, Egypt

*Abstract*—Web Application Firewalls are crucial for protecting web applications against a wide range of cyber threats. Traditional Web Application Firewalls (WAFs) often struggle to effectively distinguish between malicious and legitimate traffic, leading to limited efficacy in threat detection. To overcome these limitations, this paper proposes an Adaptive Dual-Layer WAF (ADL-WAF) employing a two-layered Machine Learning (ML) model designed to enhance the accuracy of anomaly and threat detection. The first layer employs a Decision Tree (DT) algorithm to detect anomalies by identifying traffic deviations from established normal patterns. The second layer employs Support Vector Machine (SVM) to classify these anomalies as either threat anomalies or benign anomalies. Our "Adaptive Dual-Layer WAF (ADL-WAF)" incorporates comprehensive data pre-processing and feature engineering techniques and has been thoroughly evaluated using five large benchmark datasets. Evaluation using these datasets shows that ADL-WAF achieves a detection accuracy of 99.88% and a precision of 100%, significantly enhancing anomaly detection and reducing false positives. These findings suggest that integrating machine learning techniques into WAFs can substantially improve web application security by providing more accurate and efficient threat detection.

## 1. INTRODUCTION

The widespread adoption and critical importance of web applications have made them prime targets for an increasing number of cyberattacks. Traditional Web Application Firewalls [1], however, are developed to protect these applications via inspecting and controlling HTTP traffic. They mainly use Application Learning (AL) to learn normal user behavior and identify suspicious ones. Despite their popularity among organizations, traditional WAFs exhibit several shortcomings that hinder their effectiveness. Some of these include high false positive rates, time-consuming fine-tuning processes, and a static nature in responding to changes in threats and the behavior of the applications.

The essence of the traditional WAF functionality is in the possibility of generating profiles based on traffic analysis. Nevertheless, this approach has some inherent drawbacks. Manually validating these profiles prior to deployment is both time-consuming and labor-intensive. Also, there are risks connected to the learning phases' imperfection or incorrectness: fluctuations in legitimate traffic may be recognized as threats, while real threats may be deemed benign. These are some of the problems that require web application security to be more dynamic and self-aware.

This study introduces an Adaptive Dual-Layer Web Application Firewall (ADL-WAF) that leverages machine learning to enhance the detection and categorization capabilities of traditional WAFs. The proposed system comprises two distinct layers: an Anomaly Detection Layer, which utilizes Decision Tree (DT) algorithms to identify deviations from normal traffic patterns, effectively flagging anomalous activities; and a Threat detection Layer, employing Support Vector Machines (SVM) to distinguish between benign anomalies and actual attacks, ensuring precise threat identification.

Key Contributions to This Paper:

- Introduced a novel approach for detection by developing the ADL-WAF, which combines DT and SVM algorithms to create a robust WAF capable of accurately detecting and classifying web-based threats.
- Demonstrated that the ADL-WAF outperforms traditional WAFs by reducing false positives and improving overall detection rates.
- Highlighted how the implementation of this two-layer ML framework can alleviate the workload of security teams, leading to more efficient web application protection.

The remainder of this paper is organized as follows: Section 2 reviews related work; Section 3 details the Methods; Section 4 discusses the experimental setup and results; and Section 5 presents the conclusions.

## 2. RELATED WORK

In recent years, researchers have shifted their focus towards enhancing distinguishing web attacks with the aid of various proposed models. Some studies approach conventional solutions like detecting web application attacks through attack signatures and keywords while other studies adopt machine learning-based solutions, which will be the focus of our discussion.

Zhang et al. [2] provided a framework that uses seven extracted features: Web resource, attribute sequence, attribute value, HTTP version, header, and header input value. The framework integrates three key components: probability distribution model, Hidden Markov model, and one class SVM model. All these components are trained on a dataset that only contains normal requests, and the models were evaluated on the Wikipedia access traces [3] and FuzzDB [4] datasets. This approach, which involved multiple models, improved detection accuracy by a large margin. However, the method involves feeding different request fields to different models for the detection of anomalies which may be an issue in terms of performance especially when dealing with real-time WAF services where time is of the essence.

Tekerek and Bay [5] also presented a kind of hybrid model that includes signature-based detection for known attacks and behavior-based detection for unknown attacks. They employed neural networks with three mathematically defined features as the input to the network. This model has features of hybrid detection methods, which enhance the efficiency and speed of the model; the elegance of the neural networks was also one of the reasons for the successive developments of the model. However, the extracted features are not very flexible and can be used only for web applications where the features are extracted from the logs of the web applications. Moreover, when statistical functions are used in feature extraction, it is possible to produce some of the features which are hard to handle.

Sharma et al. [6] devoted a lot of attention to how to select seven features from incoming requests while comparing the performance of three classification algorithms. They applied preprocessing techniques to the CSIC 2010 [7] dataset to address the issue of missing features, which allowed them to identify subcategories of malicious requests. While this method produced useful characteristics, the CSIC 2010 dataset had restrictions that prevented some features from being retrieved, such as the lack of cookie-length data. Furthermore, the study's dependence on a single dataset had limitations.

Hoang [8] utilized a supervised machine learning approach, specifically an inexpensive decision tree algorithm, to detect four major web attacks: Some of the common types of attacks include SQL injection, cross-site scripting, command injection and path manipulation. He did this in combination with an n-gram model for which n was set to 3 and Principal Component Analysis for dimensionality reduction. Hoang then applied the model with HTTPParams and CSIC 2010 dataset and he got an average accuracy of 98%. 56%. However, the model is developed from only the HTTPParams dataset, and only one algorithm was employed in the experiments.

Niu and Li [9] used eight statistical features of CSIC 2010, which improved detection performance with CNN and GRU. In this way, detections made by the algorithm got to 99.00% accuracy. better than other deep learning methods. However, the use of deep learning approaches for integration into real-time WAF may also make this process less efficient in terms of performance (speed), and only one dataset was used to train it.

Another notable work on the design of WAF-based model learning which focuses on using machine learning techniques and feature engineering in inspecting common web-based attacks by Aref Shaheed, M.H.D. Bassam Kurdy. [10] To do this, they start with and analyze the incoming requests to extract 4 main features: Request length Allowed Percentage Special Characters Weighted Attack Requests (this feature combines URL, Payload and headers). They tested the model with various datasets including CSIC 2010, HTTPParams 2015 [11] dataset and a new hybrid data based on both CSIC and HTTP params along with real webserver logs of compromised server- leading to an accuracy rate above 99.60% with research dataset and up-to 98.00% with Real-world. That said, there are some limitations to this approach. Whilst it is quite true, the assumption that legitimate requests are longer than malicious ones are not always valid, giving a substantial number of fallacies. Furthermore, being dependent on special characters causes some false positives; nowadays there are a variety of request-making techniques, and all these special character combinations might appear in the normal-looking application traffic too. Also, the emphasis on attack weight determined by third-party antivirus tools could abate core machine learning and feature engineering because it introduces external dependencies that might influence model consistency.

Haruta and Ryoichi. [12] offer two autoencoder-based models with an unsupervised learning approach that focuses on normal requests. The first model converts HTTP requests into ASCII codes and uses an autoencoder to learn patterns in normal requests, the second model generates word vector arrays using fastText and employs a convolutional autoencoder, for accuracy improvement. Evaluated on the HTTP dataset CSIC2010, the second model achieved an accuracy of 0.94, enhancing the first model's accuracy 0.71. However, the model is only developed using the CSIC2010 dataset, also using unsupervised learning approach not efficient all the time specially in web application anomaly detections where the anomalies are rare and need explicit identification.

While these studies also have their contribution in moving the field forward, they just differentiate between requests being normal or anomalous and most of them are using a single dataset to estimate. Some other work focuses on several types of attack as well, for example the method discussed by Dr. Ahmad Ghafarian [13], where a marker line is injected into database tables to detect SQL injections attacks. His algorithm examines the queries to be executed first and detects a malicious request with the marker line fetch. This proved to be top quality in real time SQL injection detection, however these only target one type of attack and do not consider the possible resource overhead or execution delays introduced through pre-execution query testing.

Existing machine learning-based Web Application Firewalls (WAFs) face challenges such as high false positive rates, limited adaptability to evolving threats, scalability issues, resource intensiveness, and difficulties in handling zero-day attacks. These limitations necessitate advanced solutions like the proposed Adaptive Dual-Layer WAF (ADL-WAF), which integrates machine learning techniques to enhance detection capabilities and operational efficiency.

## 3. METHODS

Traditional Web Application Firewalls (WAFs) often rely on application learning through traffic observation to identify and block malicious requests. While this method can be effective, it frequently results in high rates of false positives and false negatives. False positives occur when legitimate traffic is incorrectly flagged as malicious, leading to the unintended blocking of valid requests. Conversely, false negatives happen when actual attack patterns, especially those employing whitelisted characters to bypass detection, go unnoticed, allowing malicious activities to proceed undetected.

To address these challenges, we propose an Adaptive Dual-Layer Web Application Firewall (ADL-WAF) that integrates machine learning techniques to enhance detection accuracy and reduce false positives and negatives.

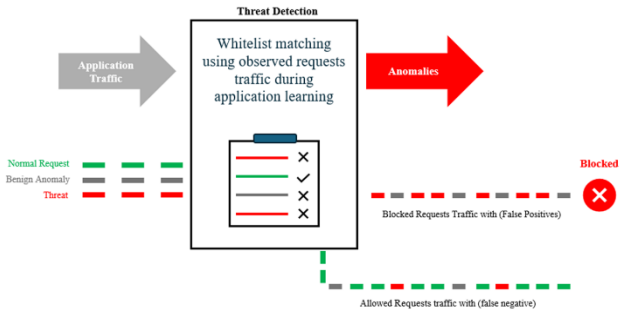

Figure 1: Traditional Web Application Firewall Build.

### 1- Proposed Model Architecture and Logic

The Adaptive Dual-Layer WAF (ADL-WAF) uses a two-layer approach, via Machine Learning (ML), that overcomes the shortcomings typically associated with traditional WAFs: Anomaly Detection (Layer 1) and Threat Detection (Layer 2). In turn, this significantly improves the accuracy and efficacy of WAF which is one with fewer false positives and false negatives, by using these layers in conjunction with each other (see Figure 2).

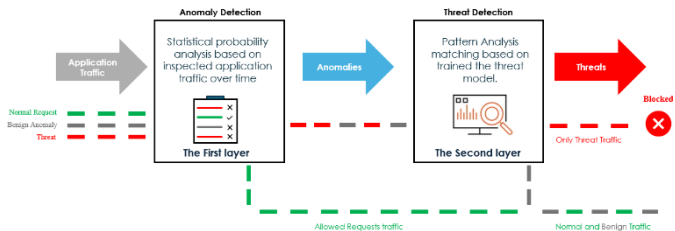

Figure 2: The Adaptive Dual-Layer WAF Build

In Fig.3 shows how The Adaptive Dual-Layer WAF's logic works and operates on different user data inputs differentiating between the legal traffic, benign inputs like human typo or error and the real threats. In the first scenario the user enters his first and last name correctly in the form field. These entries are inserted in the URL parameters and adhere to the first layer of detection. No anomaly is detected, and the user is allowed. The second scenario user mistakenly enters the character "&" which triggers an anomaly by the first layer, however the second layer checks it against the threat models and verifies it is not a threat. The third scenario is an attacker injecting SQL code into a parameter. The first layer ML identifies it as an anomaly and the second layer identifies it as an attack and it is blocked.

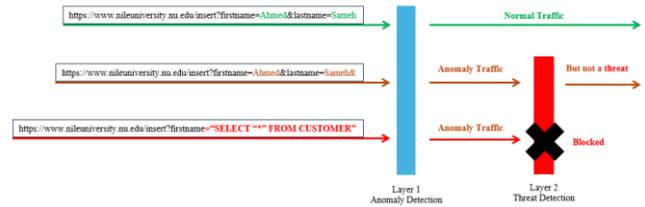

Figure 3: The Adaptive Dual-Layer WAF's decision logic.

From the previous we can determine The Adaptive Dual-Layer WAF (ADL-WAF) decision logic as follows:

- **Allow if (L1 = 0):** Normal Traffic
- **Allow if (L1 = 1) AND (L2 = 0):** Anomaly, But NOT Attack
- **Block if (L1 = 1) AND (L2 = 1):** Anomaly and Attack.

### 2- Feature engineering

#### I. Layer 1: Anomaly Detection

The first layer is responsible for the detection of anomalies in incoming web traffic. Using features extracted from HTTP requests, it identifies deviations from normal behavior. This model extracts 4 primary features simply from HTTP requests: HTTP method, absolute URL (route), body and headers, these features are then fed into calculating specific metrics on which the model classifies.

- *Alphanumeric Character Ratio:* This feature simply counts the ratio of alpha-numeric characters in a payload. Regular requests have a higher ratio of numeric and alphabetic characters compared to special symbols. As a result, this feature tends to have more value in normal requests than those from an attack.
- *Badwords Ratio:* The represents the ratio of the Number bad words (terms commonly used as parts in attack queries) to alpha-numerical character length. Normal requests have no bad words, but anomalous requests usually have a bigger ratio.
- *Special Character Ratio:* This feature quantifies the ratio of special characters (non-alphanumeric) to the total input length. Often anomalous requests have more special characters like symbols than numeric and alphabet. Therefore, this property will have a higher value in anomalous requests than normal.

- *Illegal Special Character Ratio:* This feature represents the ratio of illegal special characters to the total number of special characters in the payload. Normal requests have a low ratio or absence of illegal special characters, while anomalous requests exhibit a higher ratio.

So, the dataset that entered the ML model consists of four features with the label, label 1 for anomaly and 0 for legitimate requests. (see Table 1)

Table 1: Sample of the dataset after feature extraction

| Alphanumeric Character Ratio | Badwords Ratio | Special Character Ratio | Illegal Special Character Ratio | label |
|---|---|---|---|---|
| 84 | 0 | 15 | 0 | 0 |
| 84 | 0 | 15 | 0 | 0 |
| 85 | 0 | 14 | 0 | 0 |
| 80 | 0 | 20 | 50 | 1 |
| 50 | 0 | 50 | 50 | 1 |
| 75 | 0 | 25 | 22 | 1 |
| 85 | 0 | 14 | 0 | 0 |

### II. Layer 2: Threat Detection

It verifies whether an anomaly detected in Layer 1 is a true threat or just a benign one. We used TF-IDF (Term Frequency Inverse Document Frequency) vectorization technique [14] to get the numerical feature vector which fed into where threat detection layer does its job by identifying potential security threats like detecting path-traversal attacks, SQL injection vulnerabilities, command injection attempts and cross-site scripting. (see Table 2)

*3- Datasets*

Several datasets were utilized across different layers, we used two datasets CSIC2010, HttpParams for the anomaly detection layer, and we used three datasets HttpParams, ECML [15], XSS [16] for Threat detection layer, and unseen real-world application dataset for testing the Adaptive Dual Layer WAF (ADL-WAF) (see Table 3).

Table 3: Number of Samples in each dataset.

| Dataset Name | Total number of Samples |
|---|---|
| CSIC2010 | 61,000 |
| HTTPParams | 31,067 |
| XSS | 13,687 |
| ECML | 23,893 |
| Real Application | 30,690 |

*4- Preprocessing Datasets*

Data Preprocessing is a crucial process that converts raw data into a clean and efficient form that is suitable for our machine-learning model. Each layer has its preprocessing techniques, which depend on the model used.

The preprocessing of datasets in the anomaly detection layer consists of decoding the encoded data and then cleaning the data by removing missing values, duplicates, and outliers to ensure data integrity. After that, some initial features such as the HTTP method, absolute URL, payload, and headers are extracted from the incoming HTTP requests. Then, after extracting those features, feature selection is conducted to determine and retain only those features that are relevant to the model. In the next step, dimensionality reduction refines the dataset further to include only the most relevant features. Ultimately, balancing the dataset is done to ensure an equal distribution of normal and anomaly samples, which is crucial for the accuracy and efficiency of the model.

Likewise in the preprocessing of the datasets in the threat detection layer, the procedure starts with the conversion of the dataset from CSV to JSON format, making it compatible with the employed machine learning algorithm. After that, datasets are combined in a single file by adding rows from two or more sources, often with random shuffling to increase diversity. Then, data cleaning is done to remove missing values, duplicates, and outliers, which would ensure the quality of the dataset. After cleaning, the raw text data is changed into numeric feature vectors, a process called data vectorization. And finally, the numerical features are preprocessed to act as an input for machine learning models.

*5- Training*

In the first layer, the anomaly detection layer, the three datasets were fed into the Decision Tree (DT) classifier through two methodologies: an 80-20 train-test split and 100-fold cross-validation. To mitigate the risk of overfitting and enhance the model's generalizability [17], data rows were mixed and shuffled during training.

In the second layer, the threat detection layer, a vectorized SVM model using TFIDF (Term Frequency Inverse Document Frequency) was implemented. The efficacy of the proposed model was evaluated utilizing updated datasets, which encompassed a synthesis of various datasets. A cross-validated grid search was used to determine the best setting of hyperparameters which would yield a better performance for the model.

*6- Evaluation Metrics*

For the effective evaluation of the Adaptive Dual-Layer WAF (ADL-WAF) in identifying web attacks, we adopt three well-recognized metrics: recall, precision, and accuracy, referring to the detection rate, the false positive rate, and the overall correct decision rate achieved by the model, respectively. These metrics all rely on specific concepts of TP (True Positive), FP

(False Positive), TN (True Negative), and FN (False Negative). The detection rate (also true positive rate) is the percentage of the actual anomalies that the system detected correctly. False positive rate (also known as false alarm rate: normal requests misclassified as anomalous) is the share of normal requests classified as anomalous. Accuracy is the percentage of requests correctly classified by the system. Requests as either normal or anomalous. The metrics are defined as follows [18]:

Table 2: the numerical feature vector for the tokenized dataset using TF-IDF Technique.

| Before Tokenization | After N-gram Tokenization | Vector (Simplified) | Label |
|---|---|---|---|
| <script>alert('XSS')</script> | [<, script, alert, (, 'XSS', ), </, <script, script alert, alert (, ( 'XSS', 'XSS' ), ) </script] | [0.176, 0.477, 0.176, 0.477, 0.477, 0.477, ...] | XSS |
| SELECT * FROM users WHERE id=1 OR 1=1; | [SELECT, *, FROM, users, WHERE, id=1, OR, 1=1, ;, SELECT *, * FROM, FROM users, users WHERE, WHERE id=1, id=1 OR, OR 1=1, 1=1 ;] | [0.477, 0.477, 0.477, 0.477, 0.477, 0.477, ...] | SQLI |
| GET /index.html HTTP/1.1 | ["GET ","ET /",...,"/1.1"] | [0.000, 0.000, 0.000, 0.000, 0.000, 0.000, ...] | Normal |

- **Recall (Sensitivity or True Positive Rate):** Measures the proportion of actual threats correctly identified by the system.

$$Recall = \frac{TP}{TP + FN}$$

- **Precision (Positive Predictive Value):** Indicates the proportion of identified threats that are actual threats.

$$Precision = \frac{TP}{TP + FP}$$

- **Accuracy:** Represents the overall correctness of the system in classifying both threats and non-threats.

$$Accuracy = \frac{TP+TN}{TP+FP+FP+FN}$$

These metrics provide a comprehensive evaluation of the ADL-WAF's effectiveness in distinguishing between legitimate and malicious web traffic.

## 4. EXPERIMENTS AND RESULTS

### 1- Experimental Environment

WAF model implemented in an Anaconda environment using Python 3.1. The multi layered model has been deployed over a real-time reverse proxy setup written in Python scripting language, which listens on port 8080 to capture, process, and inspect the incoming traffic in real time.

### 2- Experiments

The objective was to enhance traditional WAFs by integrating machine learning within a two-layer framework to detect anomalies and threats. The experiments conducted include the implementation and assessment of three different machine learning models at first: Decision Tree and Naïve Bayes model for Anomaly detection and Support Vector Machine SVM for threat detection.

In the Anomaly Detection Layer, Decision Tree shows best results in the accuracy score using both train-test split and k-folds along with high sensitivity and low false positive rate, leading to its selection for our proposed model.

In the Threat Detection Layer, we used SVM with the following parameters: n-gram ranges of (1,1), (1,2), and (1,4), as well as both RBF and linear kernels, with a regularization parameter C value of 10. The best results were achieved with the RBF kernel with an n-gram range of (1,4).

Subsequently, the two layers were combined, and the proposed model logic was applied to determine the detection decision for unseen real web application samples. This integration enhanced overall accuracy by reducing false positives from 1,648 to zero, as detailed in Table 4.

Table 4: Detection metrics comparison between Anomaly Detection Layer and ADL-WAF.

| Metrics | Anomaly detection Layer | ADL - WAF |
|---|---|---|
| True Positive | 22,625 | 22,625 |
| True Negative | 7,835 | 7,835 |
| **False Positive** | **1648** | **0** |
| False Negative | 1 | 51 |

### 3- Results

We tested each layer individually to evaluate the accuracy of each. The first layer – Anomaly detection using Decision Tree (DT) showed a significant result of accuracy, 99.73% with the CSIC2010 dataset using with k-fold train method, 99.97% with the HTTPParams 2015 dataset, and 99.64% for the hybrid dataset which is a combination of HTTPParams 2015 & CSIC2010. The Second Layer – Threat Detection using a Support Vector Machine SVM achieved an accuracy of 99.9% along with significant recall and precision for different attack types (see Table 5).

Table 5: Precision and Recall in each Attack Type.

| Dataset | Precision | Recall |
|---|---|---|
| Command Injection | 0.99 | 0.98 |
| Path-Traversal | 1.0 | 0.97 |
| SQLI | 1.0 | 0.99 |
| Valid | 1.0 | 1.0 |
| XSS | 1.0 | 1.0 |

Integrating the second layer with the first reduced the false positive rate, acknowledging that not all anomalous requests are attacks, thereby increasing overall accuracy. Figure 4 illustrates the accuracy of the Anomaly detection layer compared to the combined two-Layers approach using an unseen custom dataset, along with precision and recall metrics of each.

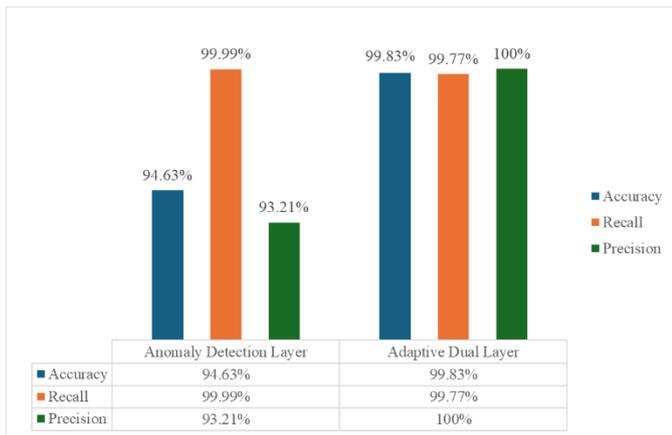

Figure 4: ADL-WAF and Anomaly detection layer comparison

## 5. DISCUSSION

The integration of SVM with TFIDF in the second layer of threat detection provides more effectiveness and adaptability to the model which helps to reduce the false positive and increase accuracy due to its high ability to detect text-numerical based attacks in web applications. This enhancement to the model makes the ADL-WAF more accurate with high precision and recall.

Notably, even prior to incorporating the second layer, the Anomaly Detection Layer demonstrated high accuracy across various datasets, outperforming related works. Table 6 presents the results for the CSIC2010, HTTPParams, and hybrid datasets, highlighting the efficacy of our preprocessing techniques and the reliability of the selected features.

Table 6: Classification accuracy of our First Layer model compared with related works.

| Related Work | CSIC | HttpParams | Hybrid Dataset |
|---|---|---|---|
| Aref, Tekerek and Bay | 99.58% | 97.61% | 96.40% |
| Sharma et al. | 96.74% | Not tested | Not tested |
| Ghafarian | 94.7% | Not tested | Not tested |
| Shaheed and Bassam | 88.32% | Not tested | Not tested |
| Ryoichi 2024 | 94% | Not tested | Not tested |
| Our Model | 99.73% | 99.97% | 99.64% |

Limitations and weaknesses of the related work Including the unreliable unigeneric features also neglecting the precision and recall metrics were taken into consideration while preparing our proposed model.

The First Layer – Anomaly detection's extracted features are reliable, dynamic and suitable for all web applications. The Use of the Dual Machine Learning model increases the overall accuracy and reduces the false positive rates which achieve an overall accuracy of 99.88% and precision of 100% for custom and unseen datasets.

While the ADL-WAF demonstrates substantial improvements, certain limitations present opportunities for future research. The current model may face challenges in identifying attacks that mimic normal requests, such as Denial of Service (DoS) and brute force attacks. Addressing this limitation would require the development of specialized, large-scale datasets over extended periods.

Implementing reinforcement learning could enable the model to adapt based on real-time feedback during decision-making processes, enhancing its dynamism and accuracy in live applications. As the model becomes more complex, ensuring that it can process incoming traffic in real-time without introducing significant latency is crucial. Future work could focus on optimizing the model's performance to maintain efficiency.

While the current feature set is effective, evaluating and enhancing the model's adaptability across a broader range of web applications with varying characteristics would be beneficial. Addressing these limitations could further enhance the ADL-WAF's robustness and applicability in diverse and evolving web security landscapes.

## 6. CONCLUSIONS

This paper introduces an innovative machine learning-based web application firewall (WAF) model, and its two-layer architecture aims to enhance anomaly and threat detection. Traditional WAFs often grapple with high false positive rates, reliance on static rule sets, and the necessity for manual configurations, limiting their responsiveness to evolving threats. The proposed Adaptive Dual-Layer WAF addresses these challenges by implementing a two-stage process of anomaly detection and a dedicated threat validity layer. The first layer looks for deviations from normal traffic patterns, and the second layer determines whether these deviations are real attacks. Using this approach effectively reduces false positives, requires less manual intervention, and obtains a more accurate traffic assessment than binary responses.

Optimizing the performance of each layer has relied on feature engineering. To detect anomalies in the first layer features such as alphanumeric ratio, special character ratio, bad words ratio, and illegal special character ratio were engineered to reflect the presence of patterns that are typical of abnormal traffic. The second layer consists of threat validation, which uses TF-IDF vectorized features to detect known attacks such as SQL Injection or Cross-Site Scripting (XSS) for instance. This layer classifies inputs using learned threat patterns by using tokenization and n-gram analysis. This dual-layer feature engineering strategy allows each layer to be thoroughly tuned for its specific task, making this model very adept at identifying and mitigating web threats, and ultimately achieving an accuracy of 99.83% on a dataset of real production environments.

Despite these advancements, certain limitations remain. The current model may encounter challenges in detecting sophisticated attacks that mimic normal traffic patterns, such as Denial of Service (DoS) and brute force attacks. Additionally, while the model demonstrates high accuracy, its adaptability to diverse web applications with varying characteristics warrants further evaluation. Future research could explore the integration of reinforcement learning to enable real-time adaptability and the development of specialized datasets to enhance detection capabilities for a broader spectrum of attack types.